\documentclass[pra,twocolumn,floatfix,superscriptaddress,longbibliography,notitlepage]{revtex4-1}

\pdfoutput=1

\usepackage[utf8]{inputenc}
\usepackage{graphicx}% Include figure files
\usepackage{natbib}
\usepackage{hyperref}% add hypertext capabilities
\usepackage{subfigure}
\usepackage{array}
\usepackage{listings}
\usepackage{pythonhighlight}
\usepackage{xcolor}

\usepackage[frozencache]{minted}
\usepackage[capitalize,nameinlink,noabbrev]{cleveref}
\usepackage{todonotes}
\usepackage{amsthm}
\theoremstyle{definition}
\newtheorem{definition}{Definition}

\theoremstyle{remark}

\theoremstyle{observation}
\newtheorem{observation}{Observation}

\newcommand{\ket}[1]{\ensuremath{\left\vert #1 \right\rangle}}

\newcommand{\qprof}{qprof}

\newcommand{\verbatimfont}[1]{\renewcommand{\verbatim@font}{\ttfamily#1}}
\newcommand{\circEqual}[2]{%
  \begin{tabular}{@{}c@{}c@{}c@{}}
    \begin{tabular}{c}
      #1
    \end{tabular}&
                   \begin{tabular}{c}
                     {\Large=}
                   \end{tabular}&
                                  \begin{tabular}{c}
                                    #2
                                  \end{tabular}%
  \end{tabular}%
}

\begin{document}

\title{\qprof{}: a gprof-inspired quantum profiler}
\author{Adrien Suau}
\email{adrien.suau@lirmm.fr}
%\orcid{0000-0002-2412-7298}
\affiliation{%
  CERFACS, 42 Avenue Gaspard Coriolis, 31057 Toulouse, France
}
\affiliation{%
  LIRMM, University of Montpellier, 161 rue Ada, 34095 Montpellier, France
}
\author{Gabriel Staffelbach}
\affiliation{%
  CERFACS, 42 Avenue Gaspard Coriolis, 31057 Toulouse, France
}
\author{Aida Todri-Sanial}
\affiliation{%
  LIRMM, University of Montpellier, CNRS, 161 rue Ada, 34095 Montpellier, France
}

\date{\today}

\begin{abstract}
  We introduce \qprof{}, a new and extensible quantum program profiler able to generate profiling reports of various quantum circuits. We describe the internal structure and working of \qprof{} and provide three practical examples on practical quantum circuits with increasing complexity. This tool will allow researchers to visualise their quantum implementation in a different way and reliably localise the bottlenecks for efficient code optimisation.
\end{abstract}

\keywords{quantum computing profiler}
\maketitle

\section{Introduction}
The quantum computing field has been evolving at an increasing rate in the past few years and is currently gaining more traction. Several quantum chips, the underlying hardware that enable researchers and companies to run quantum algorithms, have been announced by different research teams. The error rates and number of qubits provided by these chips greatly improved, with quantum hardware that have up to $65$ qubits in early 2021.

Software has also seen a tremendous rise with the emergence of several quantum computing frameworks and languages such as Qiskit \cite{Qiskit}, Q\# \cite{qsharp}, PyQuil \cite{pyquil}, Cirq \cite{cirq} or myQLM \cite{myqlm} to name a few. These frameworks help in speeding-up the process of implementing a quantum algorithm by providing their own "standard library". Most of them also include specialised libraries whose purpose is to facilitate the development and testing of new quantum algorithms. For example, all the quantum computing frameworks cited previously include a library to simulate quantum circuits, some even implement several simulation algorithms such as a full state-vector simulator, a simulator for stabilizer circuits \cite{stabilizer_Aaronson_2004,stabilizer_gidney2021stim} or a simulator using matrix-product states as described in \cite{mps_Vidal_2003,mps_Schollw_ck_2011}. Most of the frameworks that target real quantum chips also include libraries to characterise a given quantum hardware, using for example randomised benchmarking \cite{rb_Emerson_2005,rb_Knill_2008,Gambetta_2012,Cross2016,PhysRevLett.122.200502} methods, or to mitigate hardware noise \cite{mitigation_larose2020mitiq,mitigation_Bravyi_2021}.

Finally, a large majority of the quantum computing frameworks provide a way to automatically optimise a quantum circuit. This optimisation is often performed during compilation, when the abstract quantum circuit representation is translated to be compliant with the targeted hardware. Automatic optimisation of quantum circuits is a broad area of research with algorithms based on pattern-matching \cite{pattern-matching-2019arXiv190905270I,Maslov_2008,Nam2018}, gate-based optimisation algorithms \cite{gate-compilation-nn-fosel2021quantum,Bae2020} or even pulse-based ones \cite{pulse-compilation-10.1145/3297858.3304018,pulse-compilation-javadi-gokhale2020optimized, earnest2021pulseefficient}.

%Automatic optimisation of quantum circuit can be seen as the quantum analogue of what classical compilers are performing when optimising classical programs. 
But even though automatic optimisation has already been shown to be successful in optimising complex quantum circuits \cite{Childs_2018}, most algorithms only perform local optimisations without any knowledge about the algorithms used and do not have a global vision of the optimised circuit. This means that automatic optimisation algorithms will likely not be able to perform some optimisations that require a global vision of the quantum circuit or a semantic understanding of the quantum algorithms used. 
Identifying the usage of a non-optimal algorithm in the implementation and replacing it with a more efficient algorithm is an example of such an optimisation that cannot be performed without a semantic understanding of the implementation.

Currently, the only other way one has to optimise a given quantum implementation is "trial and error": try to locate a "hot spot" in the implementation, either by a tedious theoretical analysis or a manual counting of the routine calls, optimise the potential hot spot found and finally check if the optimisation performed improved the performances. This process has a severe drawback that makes it impractical on real-world implementations: the first step that consists in finding the hot spots is either imprecise or potentially very long, tedious and error-prone on large implementations.

\qprof{} aims at replacing this manual, tedious and error-prone step of analysis by automatically generating a report with all the useful information needed to find the hot spots of the given quantum program implementation. The \qprof{} tool has been strongly inspired by classical profilers such as \texttt{gprof} \cite{gprof-10.1145/872726.806987,gprof-website} which try to solve the exact same issue but in classical (non-quantum) programming. 

The rest of the paper is organised as follows. In \cref{sec:how-it-works} we explain the internals of \qprof{} and explain in details its architecture, the design choices made, and their impact on the tool efficiency, extensibility and usability. We then include code snippets and simple practical examples in \cref{sec:code-examples-and-practical-applications} to illustrate the tool usage.

\section{How does \qprof{} works?}\label{sec:how-it-works}

\subsection{General structure}\label{sec:general-structure}

The general structure of \qprof{} is composed of 3 main parts that interact with each other: framework plugins, core data structures and logic, and exporters.

\begin{figure*}
    \centering
    \includegraphics{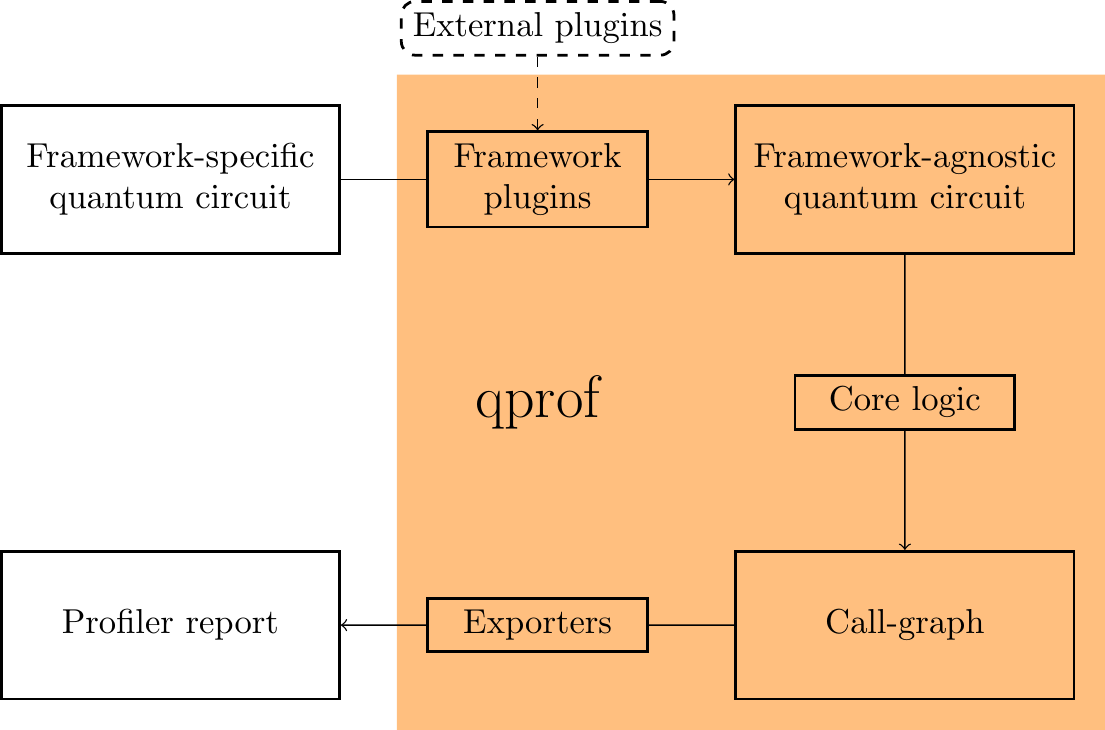}
    \caption{Schematic representation of \qprof{} workflow. Highlighted portions are part of \qprof{}.}
    \label{fig:schematic-workflow}
\end{figure*}

The overall workflow of \qprof{} is schematically explained in \cref{fig:schematic-workflow}. In this workflow, \qprof{} can be seen as a black-box that takes a "quantum circuit" as input and returns a "profiler report". This black-box view should be enough for users that only want to use the \qprof{} tool, but experienced users or plugins developers might need more details on the internals of \qprof{} in order to understand how it works.

The following sections will introduce in details the three different parts that compose \qprof{}. \Cref{sec:plugin-architecture} describes the plugins architecture, used by \qprof{} to be as modular and extensible as possible.
\Cref{sec:framework-support} then shows how this plugin architecture is used to allow \qprof{} to be mostly framework-agnostic by allowing anyone to implement a framework without modifying \qprof{} code. A description of the core data structures and core logic is then provided in \cref{sec:core-logic-and-data-structures}. Finally an explanation of the different exporters natively provided by \qprof{} is given in \cref{sec:exporters}.

\subsection{Plugin architecture\label{sec:plugin-architecture}}

The \qprof{} tool aims at being the standard for profiling quantum circuits, independently of the framework they are written with. In order to be versatile and support as much current and future quantum computing frameworks as possible, \qprof{} has been designed to be easily extensible to support new frameworks.

\qprof{} extensibility is achieved via a system of runtime-discovered plugins. In order to be discoverable by \qprof{}, a plugin should naturally obey some rules such as implementing a specific interface (explained in \cref{sec:framework-support}) or being organised in an imposed way. But the key ingredient to extensiblity is that plugins do not have to be part of the main \qprof{} tool: they can be developed and published by anyone.
This allows several situations that may help improving \qprof{} stability and evolution in time. For example, users might decide to roll-out their own plugin to support a new framework they are using internally. Another important situation that is made possible by this plugin architecture is that framework vendors have the opportunity to provide a \qprof{} plugin along with their framework and to maintain it as an official plugin.

Finally, such an architecture based on external plugins allows the user to only install the plugins and frameworks needed instead of installing all the plugins along with \qprof{}. This simple side improvement greatly reduces the installation time, installation size and plugin discovery time as it avoids installing and loading unused quantum computing frameworks.

\begin{observation}
Even though the reduction in installation time, size and loading time may seem negligible at first sight, we found out that avoiding some packages adds a significant improvement. For example, we computed that installing \texttt{qiskit} 0.24.1 with Python 3.8.8 on a modest laptop (MacBook Air from 2017) with a very good Internet connection takes more than 1 minute and ends up using more than 600 Mo of disk space. About run-time effects, importing the main \texttt{qiskit} package from a Python interpreter can take up to several seconds.
\end{observation}

\subsection{Framework support}\label{sec:framework-support}

The goal of the \qprof{} tool is to provide a unique profiling interface for quantum programs. As such, it should be able to support as much quantum computing programming frameworks as possible.
Taking into account that several of the most successful quantum computing frameworks such as Qiskit, Cirq, PyQuil or myQLM are Python libraries, and in order to ease its integration with these already existing frameworks, \qprof{} has naturally been designed as a Python library too.

But being written in the same language as a given quantum computing framework does not grant a direct compatibility. In fact, all the previously cited frameworks use a different representation for quantum programs. Qiskit implements a \texttt{QuantumCircuit} class, Cirq comes with a \texttt{Circuit} class, PyQuil uses a \texttt{Program} structure and finally \texttt{myQLM} uses either \texttt{QRoutine} or \texttt{AbstractGate} instances.

To be as independent as possible from the different quantum computing frameworks and their internal representation of a quantum circuit, \qprof{}   has been designed to be both extensible and generic. 

Extensibility is achieved with the help of the plugin architecture described in \cref{sec:plugin-architecture}.
And in order to be as generic as possible, \qprof{} uses an abstract common interface to represent the concept of "quantum (sub)routine". This concept is formally defined in \cref{def:quantum-routine,def:quantum-subroutine,def:native-quantum-subroutine}.

\begin{definition}{Quantum routine:}
    \label{def:quantum-routine}
    a (possibly parametrised) named sequence of quantum subroutines.
\end{definition}

\begin{definition}{Quantum subroutine:}
    \label{def:quantum-subroutine}
    a quantum routine that is part of a higher-level quantum routine (i.e. that is called by another quantum routine).
\end{definition}

\begin{definition}{Native quantum subroutine:}
    \label{def:native-quantum-subroutine}
    a quantum subroutine that represents a native hardware operation and that does not call any quantum subroutine.
\end{definition}

Using \cref{def:quantum-routine,def:quantum-subroutine,def:native-quantum-subroutine}, a common interface for the concept of "quantum routine" emerges. First, a quantum routine should have a name that can be retrieved. Secondly, we should be able to distinguish between native quantum subroutines and non-native ones. Finally, for non-native quantum subroutines, we need a way to iterate over all the subroutines composing it.

\begin{figure*}
    \centering
    \includegraphics[width=.9\linewidth]{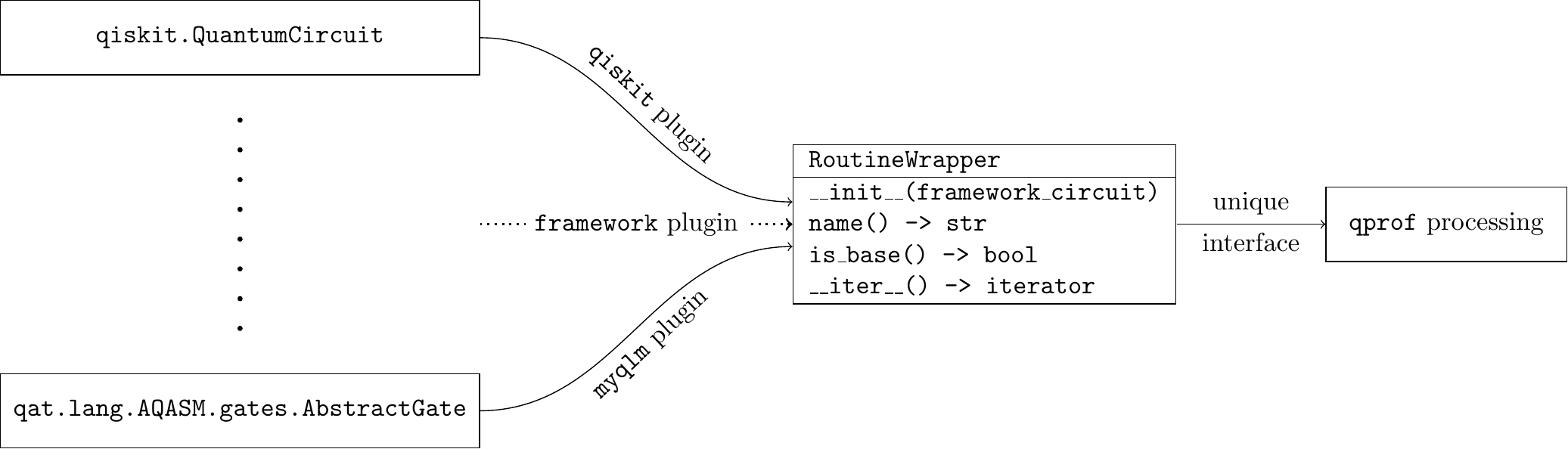}
    \caption{Schematic overview of the framework architecture used in \qprof{}. Each framework-specific representation is wrapped by a \texttt{RoutineWrapper}. Each supported framework should have a corresponding \qprof{} plugin that implements the \texttt{RoutineWrapper} interface. The \texttt{\_\_init\_\_} method initialises a \texttt{RoutineWrapper} instance with an instance of the framework-specific quantum circuit representation. The \texttt{name} method returns the name of the currently wrapped routine. \texttt{is\_base} returns \texttt{True} if the routine is a native subroutine as defined in \cref{def:native-quantum-subroutine}, else it returns \texttt{False}. Finally, the \texttt{\_\_iter\_\_} method returns an object that can be iterated on and whose iterates are the different subroutines called by the current subroutine.}
    \label{fig:schematic-framework-plugin}
\end{figure*}

This interface, schematised in \cref{fig:schematic-framework-plugin}, is the core abstraction layer of \qprof{} that allows it to be as independent as possible from the underlying quantum computing framework used to represent the profiled quantum circuit.

\subsection{Core data structures and logic}\label{sec:core-logic-and-data-structures}

Once the issue of adapting \qprof{} to the various quantum computing frameworks has been solved, we can start considering the main problem of profiling a quantum circuit. 
The first important question to answer is: what kind of figures are interesting to include in a profile report? \Cref{sec:interesting-data-to-profile} tries to address this question as extensively as possible. Then, once the target quantity to measure is known, our main concern is: how to compute this quantity from the abstract quantum circuit representation presented in \cref{sec:framework-support}? \qprof{} approach to solve this issue is presented in \cref{sec:graph-representation}.

\subsubsection{Interesting data to profile}\label{sec:interesting-data-to-profile}

Profiling a program is the action of gathering data on its execution. For classical programs and profilers, the list of data that can be gathered is quite extensive ranging from high-level quantities such as the time spent in a given function or the memory used during the program execution to low-level information recovered via hardware counters such as cache misses or branch-prediction-misses. 

But for quantum computing, the quantities of interest need to be adapted as several classical data such as cache-miss or branch-prediction-miss do not have any meaning anymore.
Nevertheless some classical quantities have a quantum analogue that may be useful for optimisation purposes. 

This is the case for the classical "instruction number" quantity, that translates trivially to its quantum counterpart "native gate number" (or "hardware gate number"). The number of native gates executed by a quantum routine is a useful information for several reasons: it is simple, the routine worst-case execution time can be computed from it and a lower-bound of the routine error rate can also be devised using this information.

Another classical quantity that can be translated to quantum computing is the "time spent in routine". This quantity can be subdivided in two more specific figures: the "time spent exclusively in routine" (sometimes called "self time") and the "time spent in subroutines called by the routine". This separation is often done in classical profiling programs as having these two execution times gives very useful information about the profiled routine that cannot be obtained from the "time spent in routine" only.

The last classical quantity with a meaningful quantum counterpart is the "memory usage", which may be translated as "number of qubits needed" when using quantum computers.

About quantities without a clear classical parallel but potentially useful, one can cite the "routine depth" as an approximation of the total execution time of the routine, the "T-count" for error-correction estimates or the "idle time" to estimate the potential effects of qubit decoherence on the routine.

\subsubsection{Graph representation (call-graph)}\label{sec:graph-representation}

Following \cref{def:quantum-routine,def:quantum-subroutine,def:native-quantum-subroutine} and the \texttt{RoutineWrapper} interface we defined in \cref{sec:framework-support}, a graph-like representation of a quantum program seems to be particularly well suited. In this representation, nodes are quantum subroutines and an oriented edge from node \texttt{A} to node \texttt{B} means that the quantum subroutine represented by \texttt{A} calls the quantum subroutine represented by \texttt{B}. This representation of a program is called a call-graph in classical computing.

\begin{figure}
    \centering
    \includegraphics[width=\linewidth]{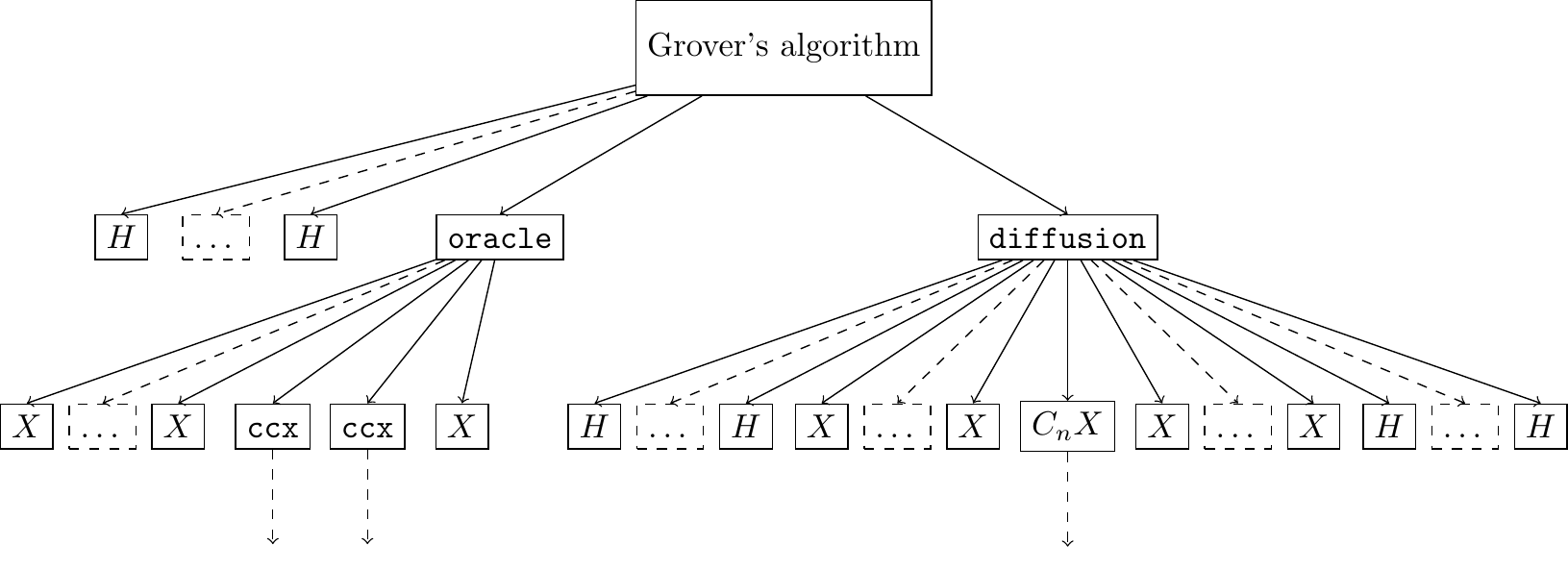}
    \caption{Call-graph representation of one possible implementation of Grover's algorithm. Dashed squares with dots within them mean a repetition of the two gates around the dashed node. Dashed arrows starting from the two \texttt{ccx} nodes and the $C_nX$ node represent a call-graph that has not been included here for readability reasons.}
    \label{fig:call-graph-duplicated-calls}
\end{figure}

Using \cref{def:quantum-routine,def:quantum-subroutine,def:native-quantum-subroutine} and the \texttt{RoutineWrapper} interface, \cref{fig:call-graph-duplicated-calls} shows a representation of one possible Grover's algorithm implementation. Even though this representation is valid according to the general definition of a call-graph, it contains a lot of redundant information that scrambles the useful data in visual noise. Because of this, most of the call graphs representations avoid the duplication of nodes, i.e. create one node for a specific subroutine and re-use this node whenever the subroutine is called.

\begin{figure}
    \centering
    \includegraphics{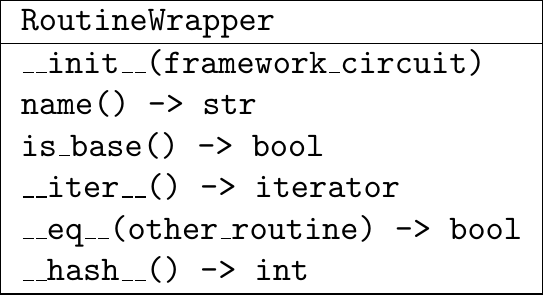}
    \caption{Final \texttt{RoutineWrapper} interface. \texttt{\_\_init\_\_}, \texttt{name}, \texttt{is\_base} and \texttt{\_\_iter\_\_} methods are described in \cref{fig:schematic-framework-plugin}. The \texttt{\_\_eq\_\_} method tests if \texttt{other\_routine} is equal to the current instance. \texttt{\_\_hash\_\_} computes an integer hash of the currently wrapper routine.}
    \label{fig:routine-wrapper-interface}
\end{figure}

And this brings to the main optimisation performed by \qprof{} under-the-hood to keep its efficiency even on very large quantum circuits: routine caching. Instead of blindly performing a full exploration of the call-graph, \qprof{} will keep in memory all the previously encountered subroutines and re-use the already computed data whenever possible. In order to make the cache efficient, \qprof{} uses a hash-map structure. This means that in order to use the cache mechanism, \texttt{RoutineWrapper} instances should now be also hashable and comparable for equality. The final \texttt{RoutineWrapper} interface that includes these requirements is represented in \cref{fig:routine-wrapper-interface}.

\begin{figure}
    \centering
    \includegraphics[width=\linewidth]{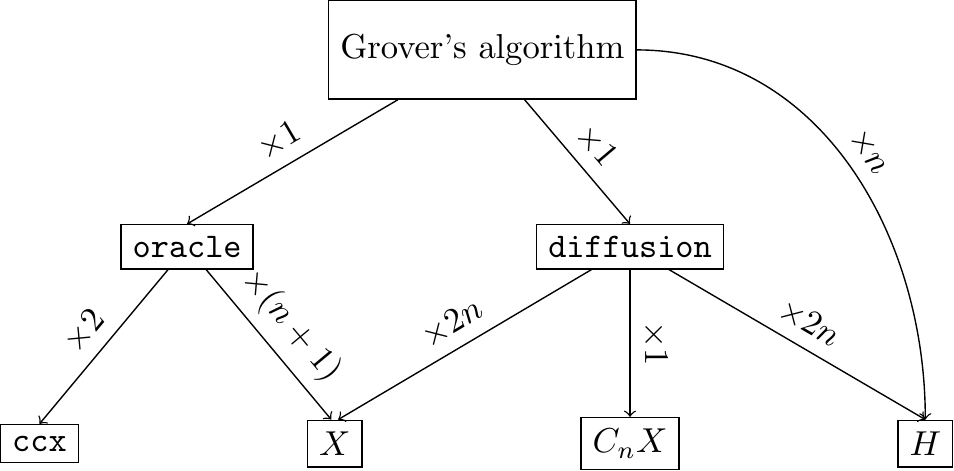}
    \caption{Call-graph representation of one possible implementation of Grover's algorithm. Each node represents one routine rather than one call to the routine. Edges have been regrouped and labelled with the number of calls for readability purposes. In the internal representation used by \qprof{}, edges are not regrouped and are ordered to account for the original quantum program subroutine call order.}
    \label{fig:call-graph-deduplicated-calls}
\end{figure}
Using such an optimisation, the call-graph represented in \cref{fig:call-graph-duplicated-calls} changes to the one in \cref{fig:call-graph-deduplicated-calls}.

It is important to realise that this optimisation does not delete any kind of information when compared to the non-optimised version. As noted in \cref{fig:call-graph-deduplicated-calls}, the order of subroutine calls is preserved and this optimisation simply allows to represent more efficiently the full call-graph.

\subsection{Exporters}\label{sec:exporters}

Along with framework support, the plugin architecture of \qprof{} is also used to implement exporters that will transform the abstract quantum program representation described in \cref{sec:graph-representation} to a more usable format. 

\begin{figure}
    \centering
    \includegraphics{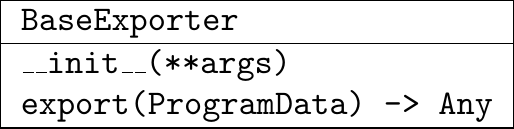}
    \caption{Exporter interface. Any plugin that implements an exporter should be a derived from the \texttt{BaseExporter} class and implement this interface. The return type of the \texttt{export} method is not specified and can be anything. The main \texttt{profile} function will return the output of the exporter.}
    \label{fig:exporter-interface}
\end{figure}

Just as the framework plugins, exporter plugins should implement a specific interface schematised in \cref{fig:exporter-interface}. \qprof{} natively implements two textual exporters: one that outputs a \texttt{gprof}-compatible format and another that returns a JSON-formatted string that directly represents a flat call-tree structure used internally by the \texttt{qprof}-compatible exporter.
%Exporters to other formats can be developed outside of \qprof{} and plugged-in effortlessly thanks to the plugin architecture presented in \cref{sec:plugin-architecture}.

\subsubsection{Flat call-tree representation}\label{sec:flat-call-tree-representation}

Before the profiler report generation, it is convenient to summarise the information contained in the generic call-graph structure presented in \cref{sec:graph-representation}. To do so, the \texttt{gprof} and JSON exporters both rely on a flat structure that represents a directed call-tree (i.e. a directed call-graph without loops).

This structure puts an additional restriction to the quantum programs that can be profiled using these exporters: the interdiction to have recursive subroutines (a subroutine that ends up calling itself).
It is important to realise that this restriction does not have a huge impact on the area of application of \qprof{} because, as of today, recursive subroutine calls are not widespread in the quantum computing programs and the restriction only applies to the \texttt{gprof} and JSON exporters, the core logic of \qprof{} being capable of handling recursive subroutine calls.

The flat call-tree structure stores, for each subroutine \texttt{A} encountered in the call-graph exploration, a list of all the subroutines \texttt{B} called by \texttt{A}. Along with each called subroutine \texttt{B}, the structure stores the number of times \texttt{B} has been called by \texttt{A} and the execution time that was needed for these calls. Finally, in order to simplify the report generation, each called routine \texttt{B} will also store a list of the routines \texttt{A} it has been called by. Within this list is also stored the number of calls to \texttt{B} that have been performed from each \texttt{A} and the execution time of these calls.

\subsubsection{\texttt{gprof} output}\label{sec:gprof-exporter-specificities}

The \texttt{gprof} exporter aims at generating a profiler report that is compatible with the profiler report returned by \texttt{gprof}, a well-known classical profiler. Being compatible with a tool that has been around for decades and is still actively used has several advantages.

First and foremost, the fact that a tool that has been stable for decades and is still actively used shows that it provides satisfaction to its users, meaning that the output format includes enough information and is sufficiently easy to read and use in practice.

Secondly, a decades-old, largely used, output format is likely to have a lot of official or user-contributed tools to help analysing and representing it in the best way possible. This is the case for the \texttt{gprof} format that can be translated to a call-graph using the \texttt{gprof2dot} tool and the \texttt{dot} executable from the Graphviz library.

Finally, the \texttt{gprof} output is simple to generate: it is a textual file with a simple and regular format.

\section{Code examples and practical applications}\label{sec:code-examples-and-practical-applications}

This section includes several examples of \qprof{} usage on various quantum circuits ranging from a simple Toffoli gate decomposition in \cref{sec:simple-program-benchmark} to more complex algorithm implementations such as Grover's algorithm in \cref{sec:grover-algorithm-benchmark}. All these benchmarks are performed on circuits generated using the \texttt{qiskit} framework. An example of benchmarking a quantum implementation of a $1$-dimensional wave equation solver written using the \texttt{myqlm} framework is finally provided in \cref{sec:pde-solver-benchmark}.

\subsection{Benchmarking a simple program}\label{sec:simple-program-benchmark}

One of the most simple quantum program that can be benchmarked is the implementation of a Toffoli gate. Such a benchmark has the benefit of being simple enough to be studied by hand which means that we will be able to verify \qprof{} results by re-computing them.

\begin{figure}
    \centering
    \resizebox{\linewidth}{!}{%
        \circEqual{%
            \includegraphics{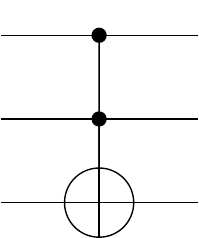}    
        }{%
            \includegraphics{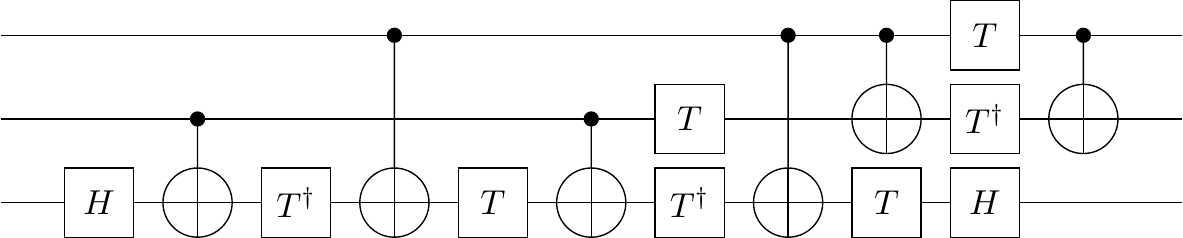}    
        }
    }
    \caption{Standard decomposition of a Toffoli gate into $1-$ and $2-$\ qubit gates.}
    \label{fig:toffoli-decomposition}
\end{figure}

The decomposition of a Toffoli gate as implemented in the \texttt{qiskit} framework is depicted in \cref{fig:toffoli-decomposition}. A complete example using \qprof{} to profile the default Toffoli gate decomposition in \texttt{qiskit} is shown in \cref{lst:toffoli-profiling}.

\begin{listing}[H]
\begin{minted}{python}
from qiskit import QuantumCircuit
from qprof import profile
# Circuit construction
circuit = QuantumCircuit(3, name="one_ccx_circuit")
circuit.ccx(0, 1, 2)
# Profiling
gate_times = {
    "u1": 0, "u2": 10, "u3": 30, "u": 30, "cx": 300
}
gprof_output = profile(circuit, gate_times, "gprof",
                       include_native_gates=True)
with open("out.qprof", "w") as f:
    f.write(gprof_output)
\end{minted}  
\caption{Python code needed to use \qprof{} on the Toffoli gate implementation and save the profiler report in a \texttt{gprof}-compatible format in a file named \texttt{out.qprof}.}
\label{lst:toffoli-profiling}
\end{listing}

\begin{figure}
    \centering
    \includegraphics[width=\linewidth]{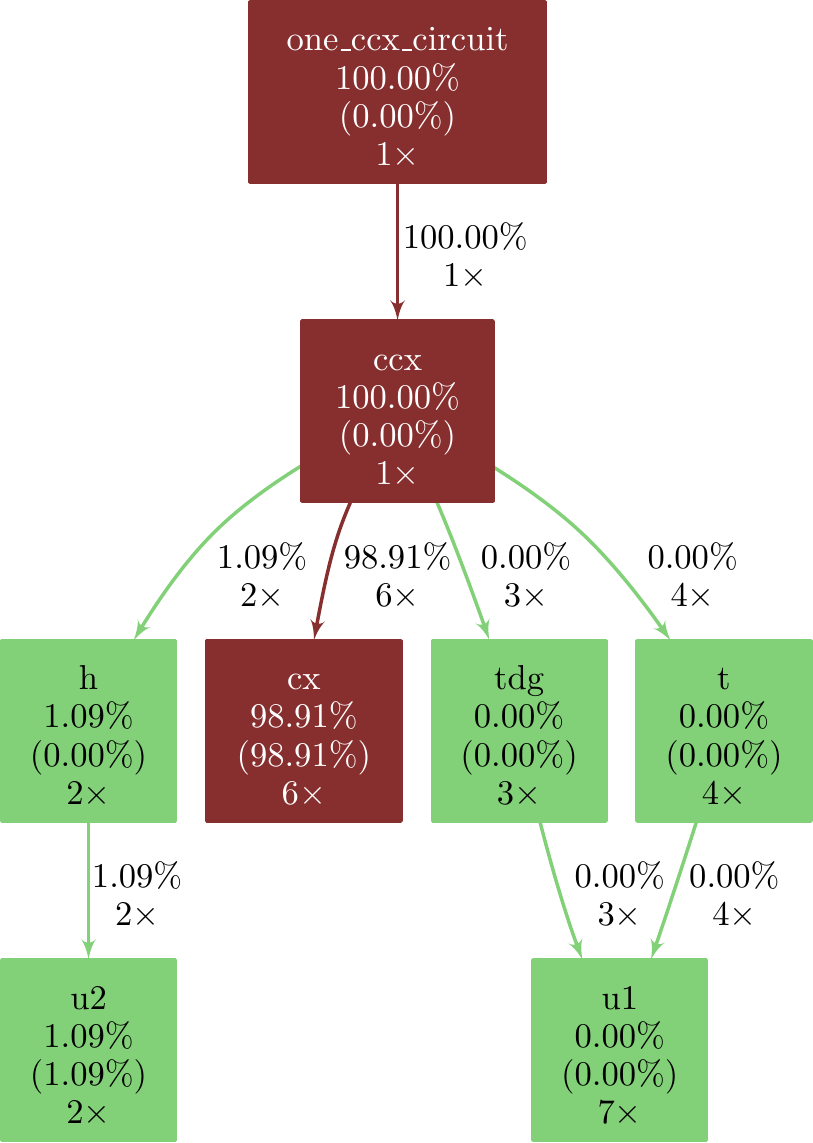}
    \caption{Call-graph for the Toffoli gate implementation. Quantum gates included in the \texttt{gate\_times} variable (here \texttt{u}, \texttt{u1}, \texttt{u2}, \texttt{u3} and \texttt{cx}) are considered as native gates. Only native gates have a non-zero self-time as they are the only gates that are really executed on the hardware.}
    \label{fig:toffoli-call-graph}
\end{figure}

The output of \qprof{}, which is here in a \texttt{gprof} format with some slight adaptations as discussed in \cref{sec:exporters}, can then be analysed. For the sake of readability and brevity, the full \texttt{gprof}-compatible profiler report will not be included verbatim in this paper and will rather be visualised using the \texttt{gprof2dot} tool that allows representing \texttt{gprof} reports as call-graphs. The call-graph obtained from the report generated in \cref{lst:toffoli-profiling} is depicted in \cref{fig:toffoli-call-graph}.

From the call-graph depicted in \cref{fig:toffoli-call-graph}, it is clear that the cost of a Toffoli gate comes from its 6 controlled-$X$ gates, that account for more than $98\%$ of the total execution time. It is also interesting to note that the $T$ gate, known to be very costly when error-correction is needed, is "free" on IBM chips when error-correction is not needed as it is equivalent to a phase change.

\subsection{Grover's algorithm}\label{sec:grover-algorithm-benchmark}

The Toffoli gate is a good example to start and understand the meaning of \qprof{}'s output but the end goal of \qprof{} is to be able to profile large and complex quantum circuits. A good first candidate to show how \qprof{} performs on a more complex circuit is Grover's algorithm.

In this example we use Grover's algorithm on four qubits to find the three quantum states that verify the following formula:
\begin{equation}
    ( q_0 \lor \neg q_1 ) \land ( \neg q_2 \land q_3 ).
    \label{eq:bool-formula}
\end{equation}
The only three $4$-qubit quantum states verifying \cref{eq:bool-formula} are \ket{0001}, \ket{1001} and \ket{1101}, $q_0$ being the left-most qubit in the bra-ket notation.

\begin{figure}
    \centering
    \includegraphics[width=\linewidth]{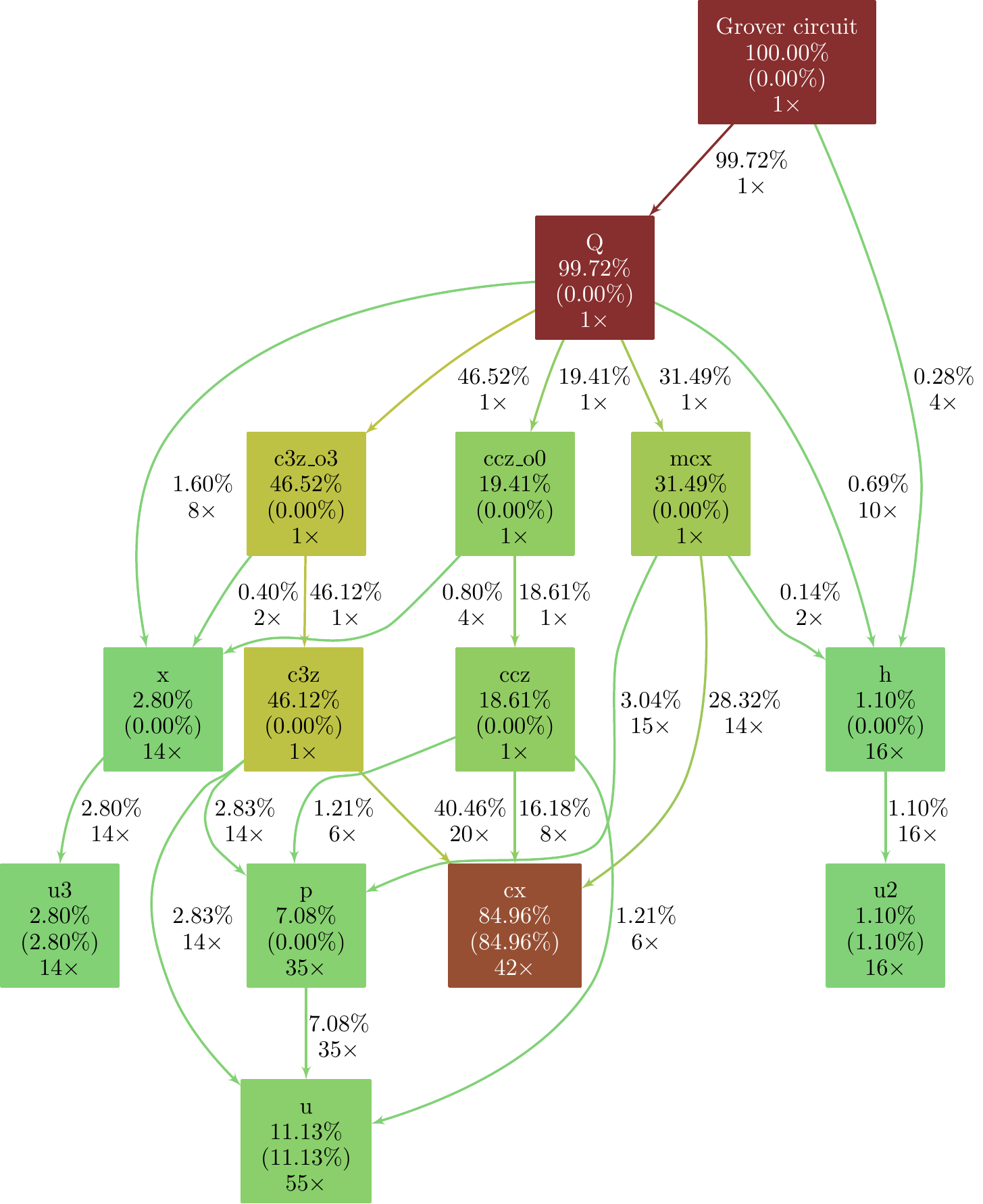}
    \caption{Call-graph for the Grover's algorithm implementation. Quantum gates included in the \texttt{gate\_times} variable (here \texttt{u}, \texttt{u1}, \texttt{u2}, \texttt{u3} and \texttt{cx}) are considered as native gates. Only native gates have a non-zero self-time as they are the only gates that are really executed on the hardware. Some percentages might not add up to exactly $100\%$ due to rounding errors.}
    \label{fig:grover-call-graph}
\end{figure}

The code needed to generate the \texttt{gprof}-compatible output for Grover's algorithm with the oracle presented in \cref{eq:bool-formula} is given in \cref{lst:grover-profiling}. The resulting call-graph, included in \cref{fig:grover-call-graph}, clearly shows that the controlled-$X$ gate is still the major contributor to execution time. But this time, contrarily to the Toffoli example shown in \cref{sec:simple-program-benchmark}, the controlled-$X$ gate is called by three different subroutines that all contribute significantly to the overall cost: \texttt{c3z}, \texttt{ccz} and \texttt{mcx}.

Thanks to \qprof{} it is now easy to understand the subroutines that contribute the most to the total execution time. More importantly, the \texttt{qprof}-compatible report and the call-graph representation gives very insightful information about subroutines calls that are crucial for circuit optimisation. Such information can be used to weight the impact of a given optimisation and then decide whether or not it is worth applying it. 

For example, knowing that the \texttt{ccz} subroutine takes $18.61\%$ of the total time, it is easy to deduce that a $20\%$ improvement in the implementation of \texttt{ccz} will translate into a tiny $\frac{18.61\%}{5} = 3.72\%$ improvement to the overall execution time, which might not be worth the effort. On the other hand, optimising the \texttt{c3z} subroutine to reduce its execution time by $20\%$ improves the overall running time by $9.22\%$, which is nearly $10\%$ and might be an interesting optimisation target. Finally, the call-graph visualisation conveys clearly the information that the \texttt{cx} gate is the most costly subroutine of the Grover's circuit, meaning that even a slight optimisation of the \texttt{cx} execution time will have a high impact on the overall implementation run-time.

\begin{listing}[H]
\begin{minted}{python}
from qiskit.algorithms import (
    AmplificationProblem, Grover
)
from qiskit.circuit.library import PhaseOracle
from qprof import profile
# Circuit construction
oracle = PhaseOracle("(v0 | ~v1) & (~v2 & v3)")
problem = AmplificationProblem(
    oracle, is_good_state=oracle.evaluate_bitstring
)
grover = Grover(iterations=1)
circuit = grover.construct_circuit(problem)
# Profile
gate_times = {
    "u1": 0, "u2": 10, "u3": 30, "u": 30, "cx": 300
}
gprof_output = profile(circuit, gate_times, "gprof",
                       include_native_gates=True)
with open("out.qprof", "w") as f:
    f.write(gprof_output)
\end{minted}
\caption{Python code needed to use \qprof{} on the Grover implementation and save the profiler report in a \texttt{gprof}-compatible format in a file named \texttt{out.qprof}.}
\label{lst:grover-profiling}
\end{listing}

\subsection{Quantum wave equation solver}\label{sec:pde-solver-benchmark}

Finally, we include in this paper a more complex example that has been implemented in a previous work with \texttt{myQLM}, a quantum computing framework maintained by Atos. The code used to generate the benchmarked quantum program is available at \url{https://gitlab.com/cerfacs/qaths/} and is explained in \cite{suau2021practical}.

\begin{figure*}
\begin{minted}{python}
from qaths.applications.wave_equation.evolve_1D_dirichlet import evolve_1d_dirichlet
from qaths.applications.wave_equation.linking_sets.arithmetic_adder import get_linking_set as arith_linking_set
from qprof import profile
# Circuit generation
time = 0.1
discretisation_size = 2 ** 10
epsilon = 1e-3
trotter_order = 1
routine = evolve_1d_dirichlet(time, discretisation_size, epsilon, trotter_order)
# Gate execution time definition
G = {"u1": 0, "u2": 89, "u3": 178, "cx": 930}
gate_times = {
    "cu1": 2 * G["cx"] + 2 * G["u1"],
    "cu2": 2 * G["cx"] + 2 * G["u3"],
    "X": G["u3"],
    "H": G["u2"],
    "CNOT": G["cx"],
    "CCNOT": 6 * G["cx"] + 2 * G["u2"] + 7 * G["u1"],
    "CH": 2 * G["cx"] + 2 * G["u3"],
    "PH": 3 * G["u1"] + 2 * G["cx"],
    "CPH": 3 * G["u1"] + 2 * G["cx"],
    "CCPH": None,
}
gate_times["CCPH"] = 3 * gate_times["CPH"] + 2 * gate_times["CCNOT"]
gate_times["CCCNOT"] = 6 * gate_times["CCNOT"] + 2 * gate_times["cu2"] + 7 * gate_times["cu1"]
# Profiling
gprof_output = profile(
    routine,
    gate_times,
    linking_set=arith_linking_set(discretisation_size),
    exporter="gprof",
)
with open("out.qprof", "w") as f:
    f.write(gprof_output)
\end{minted}  
  \caption{\normalsize Python code needed to use \qprof{} with the QatHS library, on top of myQLM, and save the profiler report in a \texttt{gprof}-compatible format in a file named \texttt{out.qprof}.}
  \label{lst:qaths-profiling}
\end{figure*}

This example demonstrates that, as can be seen in \cref{lst:qaths-profiling}, \qprof{} interface stays nearly the same even though the framework used is now completely different. The only exceptions are some additional parameters (such as \texttt{linking\_set} in \cref{lst:qaths-profiling}) that are directly forwarded to the framework plugin used and additional gate definitions in the \texttt{gate\_times} data structure because of the way gate decomposition is handled in \texttt{myQLM}.

\begin{figure}
    \centering
    \includegraphics[width=.9\linewidth, height=.9\textheight, keepaspectratio]{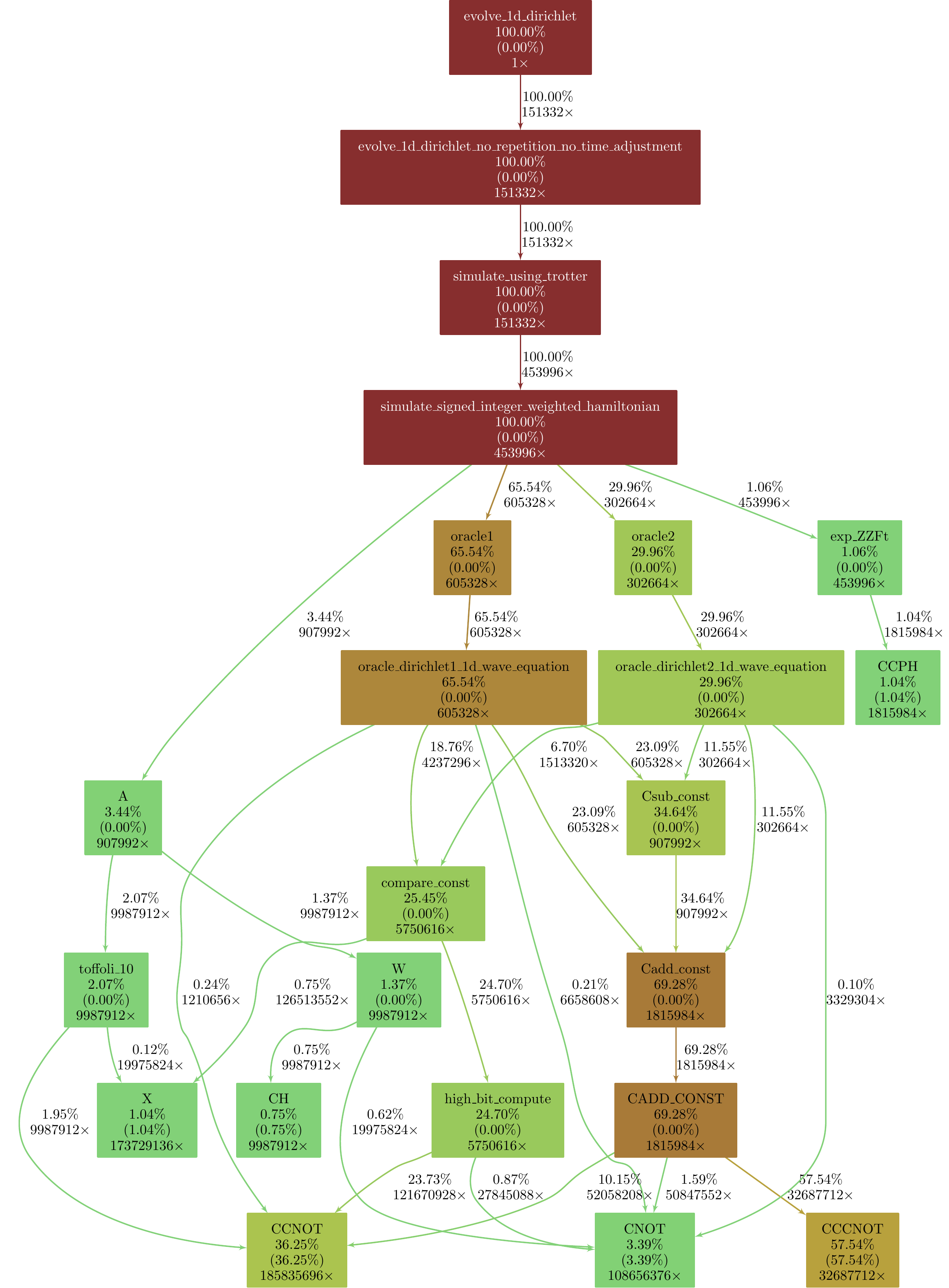}
    \caption{Call-graph of the quantum wave equation solver. Nodes (i.e. quantum routines) that account for less than $0.5\%$ of the total execution time are not plotted. Edges (i.e. subroutine calls) that account for $0.1\%$ or less of the total execution time are also discarded for readability purposes.}
    \label{fig:qaths-call-graph}
\end{figure}

The call graph obtained by running \cref{lst:qaths-profiling} is reproduced in \cref{fig:qaths-call-graph}. In order for the call-graph to be readable on a paper format, some subroutines and calls (i.e. nodes and edges respectively) have been discarded from the graphical representation. The call-graph clearly shows that most of the execution time is spent in the oracle implementation. Moreover, (multi-)controlled-$X$ gates are the major contributors to the total execution time.

\section{Conclusion}\label{sec:conclusion}

In this paper we introduced an open-source and, to the best of our knowledge, novel library that is able to generate profiling reports in well-known formats from a quantum circuit implementation. 
Our library is able to natively read quantum circuits from multiple frameworks -- currently Qiskit and myQLM -- and can be easily extended to support more quantum computing libraries. It generates consistent reports independently of the underlying framework used.
\qprof{} opens new optimisation opportunities for quantum scientists and programmers by allowing them to view their quantum circuit implementation in a well-known, familiar, synthetic and visual representation. 
In this paper, we first presented the main concepts used in the internals of \qprof{} and required to make it as extendable as possible: the plugin interface used to wrap different quantum computing frameworks, the call-graph data structure and finally the different native exporters. 
Then, we used \qprof{} on three different quantum circuit implementations of increasing complexity to demonstrate its features: simplicity of use, adaptability, consistency of the interface and efficiency.

In the future, we plan to extend the set of supported quantum computing frameworks. The number of exporters can also be improved to handle different output formats such as a \texttt{perf\_event} \cite{perf-website} compatible format, allowing to easily use new visualisations such as flame graphs \cite{flame-graph-10.1145/2909476}.

\section*{Supplementary material}
\label{sec:suppl-mater}

The \texttt{qprof} tool is available at \url{https://gitlab.com/qcomputing/qprof/qprof}.

\begin{acknowledgements}
The authors would like to acknowledge the support from TotalEnergies. The authors would like to thank Siyuan Niu for proofreading this paper.
\end{acknowledgements}

\bibliographystyle{ACM-Reference-Format}
\bibliography{bibliography}


\begin{thebibliography}{30}

%%% the Web address.

\ifx \showCODEN    \undefined \def \showCODEN     #1{\unskip}     \fi
\ifx \showDOI      \undefined \def \showDOI       #1{#1}\fi
\ifx \showISBNx    \undefined \def \showISBNx     #1{\unskip}     \fi
\ifx \showISBNxiii \undefined \def \showISBNxiii  #1{\unskip}     \fi
\ifx \showISSN     \undefined \def \showISSN      #1{\unskip}     \fi
\ifx \showLCCN     \undefined \def \showLCCN      #1{\unskip}     \fi
\ifx \shownote     \undefined \def \shownote      #1{#1}          \fi
\ifx \showarticletitle \undefined \def \showarticletitle #1{#1}   \fi
\ifx \showURL      \undefined \def \showURL       {\relax}        \fi
% invisible to TeX
\providecommand\bibfield[2]{#2}
\providecommand\bibinfo[2]{#2}
\providecommand\natexlab[1]{#1}
\providecommand\showeprint[2][]{arXiv:#2}

\bibitem[\protect\citeauthoryear{Aaronson and Gottesman}{Aaronson and
  Gottesman}{2004}]%
        {stabilizer_Aaronson_2004}
\bibfield{author}{\bibinfo{person}{Scott Aaronson} {and}
  \bibinfo{person}{Daniel Gottesman}.} \bibinfo{year}{2004}\natexlab{}.
\newblock \showarticletitle{Improved simulation of stabilizer circuits}.
\newblock \bibinfo{journal}{\emph{Physical Review A}} \bibinfo{volume}{70},
  \bibinfo{number}{5} (\bibinfo{date}{Nov} \bibinfo{year}{2004}).
\newblock
\showISSN{1094-1622}
\urldef\tempurl%
\url{https://doi.org/10.1103/physreva.70.052328}
\showDOI{\tempurl}


\bibitem[\protect\citeauthoryear{Abraham, AduOffei, Agarwal, Akhalwaya,
  Aleksandrowicz, Alexander, Amy, Arbel, Arijit02, Asfaw, Avkhadiev, Azaustre,
  AzizNgoueya, Banerjee, Bansal, et~al\mbox{.}}{Abraham et~al\mbox{.}}{2019}]%
        {Qiskit}
\bibfield{author}{\bibinfo{person}{H{\'e}ctor Abraham},
  \bibinfo{person}{AduOffei}, \bibinfo{person}{Rochisha Agarwal},
  \bibinfo{person}{Ismail~Yunus Akhalwaya}, \bibinfo{person}{Gadi
  Aleksandrowicz}, \bibinfo{person}{Thomas Alexander}, \bibinfo{person}{Matthew
  Amy}, \bibinfo{person}{Eli Arbel}, \bibinfo{person}{Arijit02},
  \bibinfo{person}{Abraham Asfaw}, \bibinfo{person}{Artur Avkhadiev},
  \bibinfo{person}{Carlos Azaustre}, \bibinfo{person}{AzizNgoueya},
  \bibinfo{person}{Abhik Banerjee}, \bibinfo{person}{Aman Bansal},
  {et~al\mbox{.}}} \bibinfo{year}{2019}\natexlab{}.
\newblock \bibinfo{title}{Qiskit: An Open-source Framework for Quantum
  Computing}.
\newblock   (\bibinfo{year}{2019}).
\newblock
\urldef\tempurl%
\url{https://doi.org/10.5281/zenodo.2562110}
\showDOI{\tempurl}


\bibitem[\protect\citeauthoryear{Bae, Alsing, Ahn, and Miller}{Bae
  et~al\mbox{.}}{2020}]%
        {Bae2020}
\bibfield{author}{\bibinfo{person}{J.-H. Bae}, \bibinfo{person}{Paul~M.
  Alsing}, \bibinfo{person}{Doyeol Ahn}, {and} \bibinfo{person}{Warner~A.
  Miller}.} \bibinfo{year}{2020}\natexlab{}.
\newblock \showarticletitle{Quantum circuit optimization using quantum Karnaugh
  map}.
\newblock \bibinfo{journal}{\emph{Scientific Reports}} \bibinfo{volume}{10},
  \bibinfo{number}{1} (\bibinfo{date}{Sept.} \bibinfo{year}{2020}).
\newblock
\urldef\tempurl%
\url{https://doi.org/10.1038/s41598-020-72469-7}
\showDOI{\tempurl}


\bibitem[\protect\citeauthoryear{Bravyi, Sheldon, Kandala, Mckay, and
  Gambetta}{Bravyi et~al\mbox{.}}{2021}]%
        {mitigation_Bravyi_2021}
\bibfield{author}{\bibinfo{person}{Sergey Bravyi}, \bibinfo{person}{Sarah
  Sheldon}, \bibinfo{person}{Abhinav Kandala}, \bibinfo{person}{David~C.
  Mckay}, {and} \bibinfo{person}{Jay~M. Gambetta}.}
  \bibinfo{year}{2021}\natexlab{}.
\newblock \showarticletitle{Mitigating measurement errors in multiqubit
  experiments}.
\newblock \bibinfo{journal}{\emph{Physical Review A}} \bibinfo{volume}{103},
  \bibinfo{number}{4} (\bibinfo{date}{Apr} \bibinfo{year}{2021}).
\newblock
\showISSN{2469-9934}
\urldef\tempurl%
\url{https://doi.org/10.1103/physreva.103.042605}
\showDOI{\tempurl}


\bibitem[\protect\citeauthoryear{Childs, Maslov, Nam, Ross, and Su}{Childs
  et~al\mbox{.}}{2018}]%
        {Childs_2018}
\bibfield{author}{\bibinfo{person}{Andrew~M. Childs}, \bibinfo{person}{Dmitri
  Maslov}, \bibinfo{person}{Yunseong Nam}, \bibinfo{person}{Neil~J. Ross},
  {and} \bibinfo{person}{Yuan Su}.} \bibinfo{year}{2018}\natexlab{}.
\newblock \showarticletitle{Toward the first quantum simulation with quantum
  speedup}.
\newblock \bibinfo{journal}{\emph{Proceedings of the National Academy of
  Sciences}} \bibinfo{volume}{115}, \bibinfo{number}{38} (\bibinfo{date}{Sep}
  \bibinfo{year}{2018}), \bibinfo{pages}{9456–9461}.
\newblock
\showISSN{1091-6490}
\urldef\tempurl%
\url{https://doi.org/10.1073/pnas.1801723115}
\showDOI{\tempurl}


\bibitem[\protect\citeauthoryear{{Cirq Developers}}{{Cirq Developers}}{2021}]%
        {cirq}
\bibfield{author}{\bibinfo{person}{{Cirq Developers}}.}
  \bibinfo{year}{2021}\natexlab{}.
\newblock \bibinfo{title}{Cirq}.
\newblock   (\bibinfo{year}{2021}).
\newblock
\urldef\tempurl%
\url{https://doi.org/10.5281/ZENODO.4062499}
\showDOI{\tempurl}


\bibitem[\protect\citeauthoryear{Computing}{Computing}{2021}]%
        {pyquil}
\bibfield{author}{\bibinfo{person}{Rigetti Computing}.}
  \bibinfo{year}{2021}\natexlab{}.
\newblock \bibinfo{title}{PyQuil documentation}.
\newblock   (\bibinfo{year}{2021}).
\newblock
\urldef\tempurl%
\url{https://pyquil-docs.rigetti.com/en/stable/}
\showURL{%
\tempurl}


\bibitem[\protect\citeauthoryear{Cross, Magesan, Bishop, Smolin, and
  Gambetta}{Cross et~al\mbox{.}}{2016}]%
        {Cross2016}
\bibfield{author}{\bibinfo{person}{Andrew~W Cross}, \bibinfo{person}{Easwar
  Magesan}, \bibinfo{person}{Lev~S Bishop}, \bibinfo{person}{John~A Smolin},
  {and} \bibinfo{person}{Jay~M Gambetta}.} \bibinfo{year}{2016}\natexlab{}.
\newblock \showarticletitle{Scalable randomised benchmarking of non-Clifford
  gates}.
\newblock \bibinfo{journal}{\emph{npj Quantum Information}}
  \bibinfo{volume}{2}, \bibinfo{number}{1} (\bibinfo{date}{April}
  \bibinfo{year}{2016}).
\newblock
\urldef\tempurl%
\url{https://doi.org/10.1038/npjqi.2016.12}
\showDOI{\tempurl}


\bibitem[\protect\citeauthoryear{Earnest, Tornow, and Egger}{Earnest
  et~al\mbox{.}}{2021}]%
        {earnest2021pulseefficient}
\bibfield{author}{\bibinfo{person}{Nathan Earnest}, \bibinfo{person}{Caroline
  Tornow}, {and} \bibinfo{person}{Daniel~J. Egger}.}
  \bibinfo{year}{2021}\natexlab{}.
\newblock \bibinfo{title}{Pulse-efficient circuit transpilation for quantum
  applications on cross-resonance-based hardware}.
\newblock   (\bibinfo{year}{2021}).
\newblock
\showeprint[arxiv]{quant-ph/2105.01063}


\bibitem[\protect\citeauthoryear{Emerson, Alicki, and Życzkowski}{Emerson
  et~al\mbox{.}}{2005}]%
        {rb_Emerson_2005}
\bibfield{author}{\bibinfo{person}{Joseph Emerson}, \bibinfo{person}{Robert
  Alicki}, {and} \bibinfo{person}{Karol Życzkowski}.}
  \bibinfo{year}{2005}\natexlab{}.
\newblock \showarticletitle{Scalable noise estimation with random unitary
  operators}.
\newblock \bibinfo{journal}{\emph{Journal of Optics B: Quantum and
  Semiclassical Optics}} \bibinfo{volume}{7}, \bibinfo{number}{10}
  (\bibinfo{date}{Sep} \bibinfo{year}{2005}), \bibinfo{pages}{S347–S352}.
\newblock
\showISSN{1741-3575}
\urldef\tempurl%
\url{https://doi.org/10.1088/1464-4266/7/10/021}
\showDOI{\tempurl}


\bibitem[\protect\citeauthoryear{Foundation}{Foundation}{2020a}]%
        {gprof-website}
\bibfield{author}{\bibinfo{person}{Free~Software Foundation}.}
  \bibinfo{year}{2020}\natexlab{a}.
\newblock \bibinfo{title}{GNU gprof}.
\newblock   (\bibinfo{year}{2020}).
\newblock
\urldef\tempurl%
\url{https://sourceware.org/binutils/docs/gprof/index.html}
\showURL{%
\tempurl}


\bibitem[\protect\citeauthoryear{Foundation}{Foundation}{2020b}]%
        {perf-website}
\bibfield{author}{\bibinfo{person}{The~Linux Foundation}.}
  \bibinfo{year}{2020}\natexlab{b}.
\newblock \bibinfo{title}{perf\_event tutorial}.
\newblock   (\bibinfo{year}{2020}).
\newblock
\urldef\tempurl%
\url{https://perf.wiki.kernel.org}
\showURL{%
\tempurl}


\bibitem[\protect\citeauthoryear{Fösel, Niu, Marquardt, and Li}{Fösel
  et~al\mbox{.}}{2021}]%
        {gate-compilation-nn-fosel2021quantum}
\bibfield{author}{\bibinfo{person}{Thomas Fösel},
  \bibinfo{person}{Murphy~Yuezhen Niu}, \bibinfo{person}{Florian Marquardt},
  {and} \bibinfo{person}{Li Li}.} \bibinfo{year}{2021}\natexlab{}.
\newblock \bibinfo{title}{Quantum circuit optimization with deep reinforcement
  learning}.
\newblock   (\bibinfo{year}{2021}).
\newblock
\showeprint[arxiv]{quant-ph/2103.07585}


\bibitem[\protect\citeauthoryear{Gambetta, Córcoles, Merkel, Johnson, Smolin,
  Chow, Ryan, Rigetti, Poletto, Ohki, and et~al.}{Gambetta
  et~al\mbox{.}}{2012}]%
        {Gambetta_2012}
\bibfield{author}{\bibinfo{person}{Jay~M. Gambetta}, \bibinfo{person}{A.~D.
  Córcoles}, \bibinfo{person}{S.~T. Merkel}, \bibinfo{person}{B.~R. Johnson},
  \bibinfo{person}{John~A. Smolin}, \bibinfo{person}{Jerry~M. Chow},
  \bibinfo{person}{Colm~A. Ryan}, \bibinfo{person}{Chad Rigetti},
  \bibinfo{person}{S. Poletto}, \bibinfo{person}{Thomas~A. Ohki}, {and}
  \bibinfo{person}{et al.}} \bibinfo{year}{2012}\natexlab{}.
\newblock \showarticletitle{Characterization of Addressability by Simultaneous
  Randomized Benchmarking}.
\newblock \bibinfo{journal}{\emph{Physical Review Letters}}
  \bibinfo{volume}{109}, \bibinfo{number}{24} (\bibinfo{date}{Dec}
  \bibinfo{year}{2012}).
\newblock
\showISSN{1079-7114}
\urldef\tempurl%
\url{https://doi.org/10.1103/physrevlett.109.240504}
\showDOI{\tempurl}


\bibitem[\protect\citeauthoryear{Gidney}{Gidney}{2021}]%
        {stabilizer_gidney2021stim}
\bibfield{author}{\bibinfo{person}{Craig Gidney}.}
  \bibinfo{year}{2021}\natexlab{}.
\newblock \bibinfo{title}{Stim: a fast stabilizer circuit simulator}.
\newblock   (\bibinfo{year}{2021}).
\newblock
\showeprint[arxiv]{quant-ph/2103.02202}


\bibitem[\protect\citeauthoryear{Gokhale, Javadi-Abhari, Earnest, Shi, and
  Chong}{Gokhale et~al\mbox{.}}{2020}]%
        {pulse-compilation-javadi-gokhale2020optimized}
\bibfield{author}{\bibinfo{person}{Pranav Gokhale}, \bibinfo{person}{Ali
  Javadi-Abhari}, \bibinfo{person}{Nathan Earnest}, \bibinfo{person}{Yunong
  Shi}, {and} \bibinfo{person}{Frederic~T. Chong}.}
  \bibinfo{year}{2020}\natexlab{}.
\newblock \bibinfo{title}{Optimized Quantum Compilation for Near-Term
  Algorithms with OpenPulse}.
\newblock   (\bibinfo{year}{2020}).
\newblock
\showeprint[arxiv]{quant-ph/2004.11205}


\bibitem[\protect\citeauthoryear{Graham, Kessler, and Mckusick}{Graham
  et~al\mbox{.}}{1982}]%
        {gprof-10.1145/872726.806987}
\bibfield{author}{\bibinfo{person}{Susan~L. Graham}, \bibinfo{person}{Peter~B.
  Kessler}, {and} \bibinfo{person}{Marshall~K. Mckusick}.}
  \bibinfo{year}{1982}\natexlab{}.
\newblock \showarticletitle{Gprof: A Call Graph Execution Profiler}.
\newblock \bibinfo{journal}{\emph{SIGPLAN Not.}} \bibinfo{volume}{17},
  \bibinfo{number}{6} (\bibinfo{date}{June} \bibinfo{year}{1982}),
  \bibinfo{pages}{120–126}.
\newblock
\showISSN{0362-1340}
\urldef\tempurl%
\url{https://doi.org/10.1145/872726.806987}
\showDOI{\tempurl}


\bibitem[\protect\citeauthoryear{Gregg}{Gregg}{2016}]%
        {flame-graph-10.1145/2909476}
\bibfield{author}{\bibinfo{person}{Brendan Gregg}.}
  \bibinfo{year}{2016}\natexlab{}.
\newblock \showarticletitle{The Flame Graph}.
\newblock \bibinfo{journal}{\emph{Commun. ACM}} \bibinfo{volume}{59},
  \bibinfo{number}{6} (\bibinfo{date}{May} \bibinfo{year}{2016}),
  \bibinfo{pages}{48–57}.
\newblock
\showISSN{0001-0782}
\urldef\tempurl%
\url{https://doi.org/10.1145/2909476}
\showDOI{\tempurl}


\bibitem[\protect\citeauthoryear{{Iten}, {Moyard}, {Metger}, {Sutter}, and
  {Woerner}}{{Iten} et~al\mbox{.}}{2019}]%
        {pattern-matching-2019arXiv190905270I}
\bibfield{author}{\bibinfo{person}{Raban {Iten}}, \bibinfo{person}{Romain
  {Moyard}}, \bibinfo{person}{Tony {Metger}}, \bibinfo{person}{David {Sutter}},
  {and} \bibinfo{person}{Stefan {Woerner}}.} \bibinfo{year}{2019}\natexlab{}.
\newblock \bibinfo{title}{Exact and practical pattern matching for quantum
  circuit optimization}.
\newblock  Article \bibinfo{articleno}{arXiv:1909.05270} (\bibinfo{date}{Sept.}
  \bibinfo{year}{2019}), \bibinfo{numpages}{arXiv:1909.05270}~pages.
\newblock
\showeprint[arxiv]{quant-ph/1909.05270}


\bibitem[\protect\citeauthoryear{Knill, Leibfried, Reichle, Britton, Blakestad,
  Jost, Langer, Ozeri, Seidelin, and Wineland}{Knill et~al\mbox{.}}{2008}]%
        {rb_Knill_2008}
\bibfield{author}{\bibinfo{person}{E. Knill}, \bibinfo{person}{D. Leibfried},
  \bibinfo{person}{R. Reichle}, \bibinfo{person}{J. Britton},
  \bibinfo{person}{R.~B. Blakestad}, \bibinfo{person}{J.~D. Jost},
  \bibinfo{person}{C. Langer}, \bibinfo{person}{R. Ozeri}, \bibinfo{person}{S.
  Seidelin}, {and} \bibinfo{person}{D.~J. Wineland}.}
  \bibinfo{year}{2008}\natexlab{}.
\newblock \showarticletitle{Randomized benchmarking of quantum gates}.
\newblock \bibinfo{journal}{\emph{Physical Review A}} \bibinfo{volume}{77},
  \bibinfo{number}{1} (\bibinfo{date}{Jan} \bibinfo{year}{2008}).
\newblock
\showISSN{1094-1622}
\urldef\tempurl%
\url{https://doi.org/10.1103/physreva.77.012307}
\showDOI{\tempurl}


\bibitem[\protect\citeauthoryear{LaRose, Mari, Karalekas, Shammah, and
  Zeng}{LaRose et~al\mbox{.}}{2020}]%
        {mitigation_larose2020mitiq}
\bibfield{author}{\bibinfo{person}{Ryan LaRose}, \bibinfo{person}{Andrea Mari},
  \bibinfo{person}{Peter~J. Karalekas}, \bibinfo{person}{Nathan Shammah}, {and}
  \bibinfo{person}{William~J. Zeng}.} \bibinfo{year}{2020}\natexlab{}.
\newblock \bibinfo{title}{Mitiq: A software package for error mitigation on
  noisy quantum computers}.
\newblock   (\bibinfo{year}{2020}).
\newblock
\showeprint[arxiv]{quant-ph/2009.04417}


\bibitem[\protect\citeauthoryear{Maslov, Dueck, Miller, and Negrevergne}{Maslov
  et~al\mbox{.}}{2008}]%
        {Maslov_2008}
\bibfield{author}{\bibinfo{person}{D. Maslov}, \bibinfo{person}{G.W. Dueck},
  \bibinfo{person}{D.M. Miller}, {and} \bibinfo{person}{C. Negrevergne}.}
  \bibinfo{year}{2008}\natexlab{}.
\newblock \showarticletitle{Quantum Circuit Simplification and Level
  Compaction}.
\newblock \bibinfo{journal}{\emph{IEEE Transactions on Computer-Aided Design of
  Integrated Circuits and Systems}} \bibinfo{volume}{27}, \bibinfo{number}{3}
  (\bibinfo{date}{Mar} \bibinfo{year}{2008}), \bibinfo{pages}{436–444}.
\newblock
\showISSN{1937-4151}
\urldef\tempurl%
\url{https://doi.org/10.1109/tcad.2007.911334}
\showDOI{\tempurl}


\bibitem[\protect\citeauthoryear{McKay, Sheldon, Smolin, Chow, and
  Gambetta}{McKay et~al\mbox{.}}{2019}]%
        {PhysRevLett.122.200502}
\bibfield{author}{\bibinfo{person}{David~C. McKay}, \bibinfo{person}{Sarah
  Sheldon}, \bibinfo{person}{John~A. Smolin}, \bibinfo{person}{Jerry~M. Chow},
  {and} \bibinfo{person}{Jay~M. Gambetta}.} \bibinfo{year}{2019}\natexlab{}.
\newblock \showarticletitle{Three-Qubit Randomized Benchmarking}.
\newblock \bibinfo{journal}{\emph{Phys. Rev. Lett.}}  \bibinfo{volume}{122}
  (\bibinfo{date}{May} \bibinfo{year}{2019}), \bibinfo{pages}{200502}.
\newblock
Issue 20.
\urldef\tempurl%
\url{https://doi.org/10.1103/PhysRevLett.122.200502}
\showDOI{\tempurl}


\bibitem[\protect\citeauthoryear{Nam, Ross, Su, Childs, and Maslov}{Nam
  et~al\mbox{.}}{2018}]%
        {Nam2018}
\bibfield{author}{\bibinfo{person}{Yunseong Nam}, \bibinfo{person}{Neil~J.
  Ross}, \bibinfo{person}{Yuan Su}, \bibinfo{person}{Andrew~M. Childs}, {and}
  \bibinfo{person}{Dmitri Maslov}.} \bibinfo{year}{2018}\natexlab{}.
\newblock \showarticletitle{Automated optimization of large quantum circuits
  with continuous parameters}.
\newblock \bibinfo{journal}{\emph{npj Quantum Information}}
  \bibinfo{volume}{4}, \bibinfo{number}{1} (\bibinfo{date}{May}
  \bibinfo{year}{2018}).
\newblock
\urldef\tempurl%
\url{https://doi.org/10.1038/s41534-018-0072-4}
\showDOI{\tempurl}


\bibitem[\protect\citeauthoryear{Schollwöck}{Schollwöck}{2011}]%
        {mps_Schollw_ck_2011}
\bibfield{author}{\bibinfo{person}{Ulrich Schollwöck}.}
  \bibinfo{year}{2011}\natexlab{}.
\newblock \showarticletitle{The density-matrix renormalization group in the age
  of matrix product states}.
\newblock \bibinfo{journal}{\emph{Annals of Physics}} \bibinfo{volume}{326},
  \bibinfo{number}{1} (\bibinfo{date}{Jan} \bibinfo{year}{2011}),
  \bibinfo{pages}{96–192}.
\newblock
\showISSN{0003-4916}
\urldef\tempurl%
\url{https://doi.org/10.1016/j.aop.2010.09.012}
\showDOI{\tempurl}


\bibitem[\protect\citeauthoryear{Shi, Leung, Gokhale, Rossi, Schuster,
  Hoffmann, and Chong}{Shi et~al\mbox{.}}{2019}]%
        {pulse-compilation-10.1145/3297858.3304018}
\bibfield{author}{\bibinfo{person}{Yunong Shi}, \bibinfo{person}{Nelson Leung},
  \bibinfo{person}{Pranav Gokhale}, \bibinfo{person}{Zane Rossi},
  \bibinfo{person}{David~I. Schuster}, \bibinfo{person}{Henry Hoffmann}, {and}
  \bibinfo{person}{Frederic~T. Chong}.} \bibinfo{year}{2019}\natexlab{}.
\newblock \showarticletitle{Optimized Compilation of Aggregated Instructions
  for Realistic Quantum Computers}. In \bibinfo{booktitle}{\emph{Proceedings of
  the Twenty-Fourth International Conference on Architectural Support for
  Programming Languages and Operating Systems}} \emph{(\bibinfo{series}{ASPLOS
  '19})}. \bibinfo{publisher}{Association for Computing Machinery},
  \bibinfo{address}{New York, NY, USA}, \bibinfo{pages}{1031–1044}.
\newblock
\showISBNx{9781450362405}
\urldef\tempurl%
\url{https://doi.org/10.1145/3297858.3304018}
\showDOI{\tempurl}


\bibitem[\protect\citeauthoryear{Suau, Staffelbach, and Calandra}{Suau
  et~al\mbox{.}}{2021}]%
        {suau2021practical}
\bibfield{author}{\bibinfo{person}{Adrien Suau}, \bibinfo{person}{Gabriel
  Staffelbach}, {and} \bibinfo{person}{Henri Calandra}.}
  \bibinfo{year}{2021}\natexlab{}.
\newblock \showarticletitle{Practical Quantum Computing: Solving the Wave
  Equation Using a Quantum Approach}.
\newblock \bibinfo{journal}{\emph{ACM Transactions on Quantum Computing}}
  \bibinfo{volume}{2}, \bibinfo{number}{1}, Article \bibinfo{articleno}{2}
  (\bibinfo{date}{2} \bibinfo{year}{2021}), \bibinfo{numpages}{35}~pages.
\newblock
\showISSN{2643-6809}
\urldef\tempurl%
\url{https://doi.org/10.1145/3430030}
\showDOI{\tempurl}
\showeprint[arxiv]{quant-ph/2003.12458}


\bibitem[\protect\citeauthoryear{team}{team}{2021a}]%
        {myqlm}
\bibfield{author}{\bibinfo{person}{Atos Quantum~Computing team}.}
  \bibinfo{year}{2021}\natexlab{a}.
\newblock \bibinfo{title}{myQLM documentation}.
\newblock   (\bibinfo{year}{2021}).
\newblock
\urldef\tempurl%
\url{https://myqlm.github.io/}
\showURL{%
\tempurl}


\bibitem[\protect\citeauthoryear{team}{team}{2021b}]%
        {qsharp}
\bibfield{author}{\bibinfo{person}{Microsoft~Quantum team}.}
  \bibinfo{year}{2021}\natexlab{b}.
\newblock \bibinfo{title}{The Q\# User Guide}.
\newblock   (\bibinfo{year}{2021}).
\newblock
\urldef\tempurl%
\url{https://docs.microsoft.com/en-us/azure/quantum/user-guide/}
\showURL{%
\tempurl}


\bibitem[\protect\citeauthoryear{Vidal}{Vidal}{2003}]%
        {mps_Vidal_2003}
\bibfield{author}{\bibinfo{person}{Guifré Vidal}.}
  \bibinfo{year}{2003}\natexlab{}.
\newblock \showarticletitle{Efficient Classical Simulation of Slightly
  Entangled Quantum Computations}.
\newblock \bibinfo{journal}{\emph{Physical Review Letters}}
  \bibinfo{volume}{91}, \bibinfo{number}{14} (\bibinfo{date}{Oct}
  \bibinfo{year}{2003}).
\newblock
\showISSN{1079-7114}
\urldef\tempurl%
\url{https://doi.org/10.1103/physrevlett.91.147902}
\showDOI{\tempurl}


\end{thebibliography}

\end{document}